\documentclass[9pt]{proto}

\usepackage[frozencache=true,cachedir=.]{minted}
\usepackage{lipsum} 
\usepackage[version=4]{mhchem}
\usepackage{siunitx}
\DeclareSIUnit\Molar{M}

\title{Characterising mosquito biting behaviour at high resolution}

\author[1,2]{Gregory PD Murray}
\author[1,2]{Emilie Giraud}
\author[1,2]{Felix JH Hol\textsuperscript{*}}

\affil[1]{Center for research and Interdisciplinarity, Universit\'{e} de Paris, Paris, France}
\affil[2]{Insect-Virus Interactions Unit, Institut Pasteur, UMR2000, CNRS, Paris, France}

\corr{felix.hol@pasteur.fr}{FJHH}

\begin{document}

\maketitle

\begin{abstract}
Blood feeding represents a critical event in the life cycle of female mosquitoes. In addition to providing nutrients to the mosquito, blood feeding facilitates the transmission of parasites and viruses to hosts, potentially having devastating health consequences. Despite this, our understanding of these short, yet important bouts of behaviour is incomplete. How and where a mosquito decides to feed and the success of feeding can influence the transmission of pathogens, while a more thorough understanding may allow interventions to reduce or prevent infections. Recent advances in machine vision and automated tracking presents the opportunity to observe and understand blood feeding behaviour of mosquitoes at unprecedented spatial and temporal resolution. Here, we combine these technologies with novel designs for behavioural arenas and controllable artificial host cues to enable detailed observations of biting behaviour using relatively inexpensive and readily available materials. Combined with a workflow for quantitative image analysis, we are able to describe nuanced, high resolution biting behaviour under tightly controlled conditions.
\end{abstract}

\section{Introduction}
The bite of a mosquito is one of the most critical steps in the transmission of mosquito-borne pathogens \citep{Takken2013} The bite, and subsequent blood meal, are preceded by a complex behavioural trajectory consisting of flying towards and landing on a host, selecting a bite site, and engorging a full blood meal \citep{Carde2015}. A great deal of research effort has been put into understanding the behaviour of mosquitoes during host-seeking flight, and when interacting with vector control tools such as bed nets and traps \citep{Cribellier2018,Murray2020}. However, a critical stage for both the reproductive success of the mosquito and the transmission of pathogens is during the act of blood feeding and associated movement on the host, with this being a high risk, infrequent event in the mosquito life cycle \citep{Walker1985,reid2014bite,Vinauger2018}. Despite the vital role blood feeding plays in the lifecycle of the mosquito and the epidemiological implications to their hosts, there are very few methods to monitor and record blood feeding behaviour at high spatial and temporal resolution. While this remains a challenging task, on-going improvements in imaging technology and computational approaches are enabling the development of novel methods to algorithmically characterize mosquito blood feeding at high spatiotemporal resolution \citep{Hol2020}. 

To record the biting behaviour of mosquitoes, several discrete elements must come together: a lure to induce landing behaviour, a bite substrate which mosquitoes will engorge on, an arena to contain the mosquitoes under experimental conditions, and a method to record and track the activity of the mosquitoes. The majority of studies quantifying blood feeding behaviour have relied on a human or rodent to act as both lure and blood source \citep{Hughes2020,choumet2012visualizing}. This practice, however, constrains the type of measurements that can be made (e.g. providing limited opportunity for imaging), may introduce variability between hosts/replicates, limits possibilities to modify the skin surface, and requires behavioural arenas to be designed around access to the host subject. In addition to these experimental limitations, previous methods often relied on the direct observation and recording of blood feeding behaviour by a human observer, limiting throughput and preventing automated analysis. More recently event logging software (e.g. BORIS by \cite{friard2016boris}) has streamlined and digitised this process, however, these approaches still require a significant time-investment due to their requirements for manually scoring behaviour, and may suffer from biases of the observer. 

To date, there have been few examples of high-resolution recording and in-depth analysis of mosquito feeding behaviour. Intravital video microscopy has been used to describe the sequence of events taking place during the feeding of \emph{Anopheles gambiae} on anesthetised mice \citep{choumet2012visualizing}. These striking videos revealed that the mouthparts of biting mosquitoes are highly mobile during probing, prior to feeding. A more recent study used videography to examine the late stage host-seeking behaviour of \emph{Anopheles gambiae} \citep{Hughes2020}. By filming mosquitoes in a 10 cm\textsuperscript{3} behavioural arena, the behaviour of the mosquitoes as they sought a blood meal from a human host was captured and the impact of insecticidal nets examined. Analysis in both the \cite{choumet2012visualizing} and \cite{Hughes2020} studies relied on manually scoring and analyzing events using real-time playback.

The relatively low costs of video capture equipment means that recording freely behaving animals is now practical with exceptional levels of accuracy. Advances in computational methods furthermore enable automated analysis and extraction of behavioural metrics from video data with sufficient efficiency to produce replicates for hypothesis testing and robust analysis of even subtle behavioural phenomena \citep{Mathis2018}. A third development is the use of open hardware tools to construct behavioural arenas tailored to address specific experimental questions. The availability of low-cost microcomputers (e.g. the Raspberry Pi), sensors, 3D printers, and other fabrication tools enables researchers to create experimental environments designed to mimic key aspects of the natural world, and record the behaviour of animals exploring these environments. These cross-disciplinary developments now enable the characterisation of a variety of mosquito behaviours at unprecedented levels of detail. 

A recent method taking advantage of these developments to record the biting behaviour of mosquitoes is the biteOscope: an open platform which makes use of a transparent behavioural arena and artificial bite substrate to allow the recording and automated detection of mosquitoes during landing, movement, and feeding \citep{Hol2020}. The biteOscope combines high-resolution imaging of mosquito probing and feeding with a machine vision workflow to extract detailed behavioural information on the locomotion, pose, biting, and feeding dynamics of the subject mosquitoes.

Below we outline several aspects that are critical to an experiment characterising blood feeding behaviour through quantitative imaging:

\subsection{Bite substrate and cage} 
In order to observe feeding behaviour, sufficient cues need to be present to induce feeding. Traditionally a live host has been used in blood feeding studies, yet more recent studies have used artificial bite substrates. While artificial substrates may present a less rich repertoire of stimuli compared to a live host, they offer several advantages: artificial substrates can be tailored to present a user-defined set of cues, can be designed to facilitate imaging, and increase reproducibility. In the absence of a live host, carbon dioxide is the most common stimulus to activate mosquitoes and initiate host seeking behaviour, and together with heat forms a minimal set of cues to induce a host seeking and probing response. Depending on the experimental design, additional cues, including odorants and visual stimuli may be presented. The actual blood meal can come directly from a host, yet biting and engorgement can also be promoted using an artificial meal. Previous work demonstrated a robust feeding response on a solution containing 1mM adenosine triphosphate (ATP), sodium ions, and an osmotic pressure close to that of blood \citep{galun1963feeding,Duvall2019}. The use of an artificial meal has the benefit of being transparent, allowing visualisation of the mosquito probing and interacting with the feeding liquid, and quantifying feeding activity. 

The bite substrate is the central component of a behavioural arena which contains the mosquitoes and should be sufficiently large to allow the mosquitoes to interact with the bite substrate. Mounting a transparent bite substrate in the floor,  wall,  or ceiling of a transparent cage has the advantage that the camera and other equipment can be located outside the  cage, while the mosquitoes can access the substrate from the inside. Laser cutting sheets  of acrylic provides a versatile and easy way to construct custom cages suited for behavioural experiments. 

\subsection{Imaging considerations}
When choosing an imaging setup there are several trade-offs to consider. Choosing the right camera and lens combination depends on the level of detail that needs to be resolved (more detail, smaller field of view) and the number of individuals to be imaged (more individuals, larger field of view). Another trade-off exists between camera resolution and frame rate against file size and the associated storage space and processing time. Striking the right balance depends on the experimental question to be addressed, for example: Is it necessary to resolve small body parts such as the labrum? What are the fastest dynamics to be characterised? Our current studies show that imaging an 8 x 8 cm bite substrate using a 12 megapixel camera (1/1.7'' sensor) and a 35 mm c-mount lens provides an image quality enabling the reliable detection of small body parts like legs and proboscis, while allowing up to 20 mosquitoes to simultaneously interact with the bite substrate. Acquiring frames at a minimum of 25 frames per second (fps) is required to temporally resolve distinct behaviours such as biting, grooming, and walking (the stride frequency of mosquitoes is approximately 10 Hz).

Correct illumination is critical to obtaining high quality images and consequently the amount of information that can be extracted from those images. LED arrays are a cost-effective means to achieve even illumination having consistent intensity and well-defined spectral properties. Experiments with day-biting mosquitoes (e.g. \emph{Aedes}) are usually conducted using visible wavelengths (e.g. white light), whereas nocturnal or crepuscular species, such as \emph{Anopheles}, are usually recorded in dark conditions using infrared (IR) illumination. Previous work has demonstrated an inability to detect IR light in both \emph{Anopheles} \citep{gibson1995behavioural} and \emph{Aedes} \citep{Muir1992} mosquitoes. 

\subsection{Recording and detection of behaviour}
Traditionally, mosquito behaviour has been characterised by either direct observation or watching playback of recorded images, possibly assisted by event logging software to manually log behaviours based on predefined classes. An alternative to the labour intensive and observer-based detection is to use machine vision for the detection and tracking of mosquito movement \citep{VanBreugel2015, angarita2016novel, Hol2020}. Automated image analysis methods may be based on the use of markers to detect and track individuals, yet recent advances in deep learning are enabling markerless tracking and pose estimation of multiple individuals in a single field of view. A recent study characterising mosquitoes interacting with an artificial bite substrate used the DeepLabCut framework \citep{Mathis2018} for the markerless tracking of the movement of all legs, head, proboscis, and abdomen of a variety of mosquito species \citep{Hol2020}. The output of body part tracking algorithms can be used to obtain behavioural statistics (e.g. locomotion, activity, feeding status) and form the basis of downstream computational pipelines, including behavioural classification. Machine learning tools to automate the extraction of behavioural statistics from large quantities of video data enables high-throughput behavioural characterization and removes biases associated with behavioural classification by human observers.

\clearpage

\section{{\Huge Protocol: Characterising mosquito biting\\
behaviour using the bite\textcolor{red}{O}scope}}
\vspace{0.4cm}
\textbf{Gregory PD Murray\textsuperscript{1,2}, Emilie Giraud\textsuperscript{1,2}, Felix JH Hol\textsuperscript{*,1,2}}\\
\textsuperscript{1}Center for research and Interdisciplinarity, Universit\'{e} de Paris, Paris, France\\
\textsuperscript{2}Insect-Virus Interactions Unit, Institut Pasteur, UMR2000, CNRS, Paris, France\\
\textsuperscript{*}felix.hol@pasteur.fr

\begin{abstract}
The biteOscope enables the high resolution monitoring and video recording of mosquito movement and behaviour during blood feeding. Mosquito biting is induced by combining an artificial blood-meal, membrane, and transparent heater in a transparent behavioural arena. Machine vision techniques enable the tracking and pose-estimation of individual mosquitoes to discern behaviour and resolve individual feeding events. The workflow allows multiple replicates and large amounts of imaging data to be generated rapidly. The obtained data is suitable for downstream analysis using machine learning tools for behavioural analysis allowing subtle behavioural effects to be characterised.
\end{abstract}

\section{Materials}

\subsection{Equipment}
\vspace{0.1cm}
\begin{itemize} 
\item Behavioural arena
\subitem BiteOscope cage made from 3mm thick clear cast acrylic (instructions below)
\subitem 250 mL tissue culture flask (Falcon flask, Thermo Fisher Scientific \# 353136)
\item Imaging setup 1
\subitem USB3 camera with c-mount (e.g. Basler acA4024-29um (4024 x 3036 pix, 31 fps))
\subitem 35 mm c-mount lens (e.g. Thorlabs MVL35M23)
\subitem Computer with USB3 port (e.g. Intel NUC with i7 processor)
\item Imaging setup 2 (low cost alternative)
\subitem Raspberry Pi HQ camera
\subitem 16 mm c-mount lens
\subitem NVIDIA  Jetson nano
\item Illumination:
\subitem White light LED array (e.g. VidPro LED 312)
\subitem \emph{or:}
\subitem 850 nm IR LED array (Taobao)
\item Temperature controller:
\subitem Two Peltier heating elements (e.g. 12V 5A Peltier, Adafruit 1330)
\subitem temperature sensor DS18b20
\subitem Raspberry Pi
\subitem USB charger (2x)
\subitem USB cable male - male, cut in half
\subitem relay switch
\subitem jumper wires
\subitem alligator clips
\end{itemize}

\subsection{Reagents}
\vspace{0.1cm}
\begin{itemize}
\item Artificial meal: 110mM NaCl, 20mM NaHCO3, and 1 mM ATP
\end{itemize}

\subsection{Software}
\vspace{0.1cm}
Code for the temperature controller and all data processing code is written in Python 3 and is available on \href{https://github.com/felixhol/biteOscope}{Github}. \href{https://github.com/DeepLabCut}{DeepLabCut} processing has been tested with version 2.2b8. 

\section{Constructing the biteOscope}

\begin{figure}
\includegraphics[width=\linewidth]{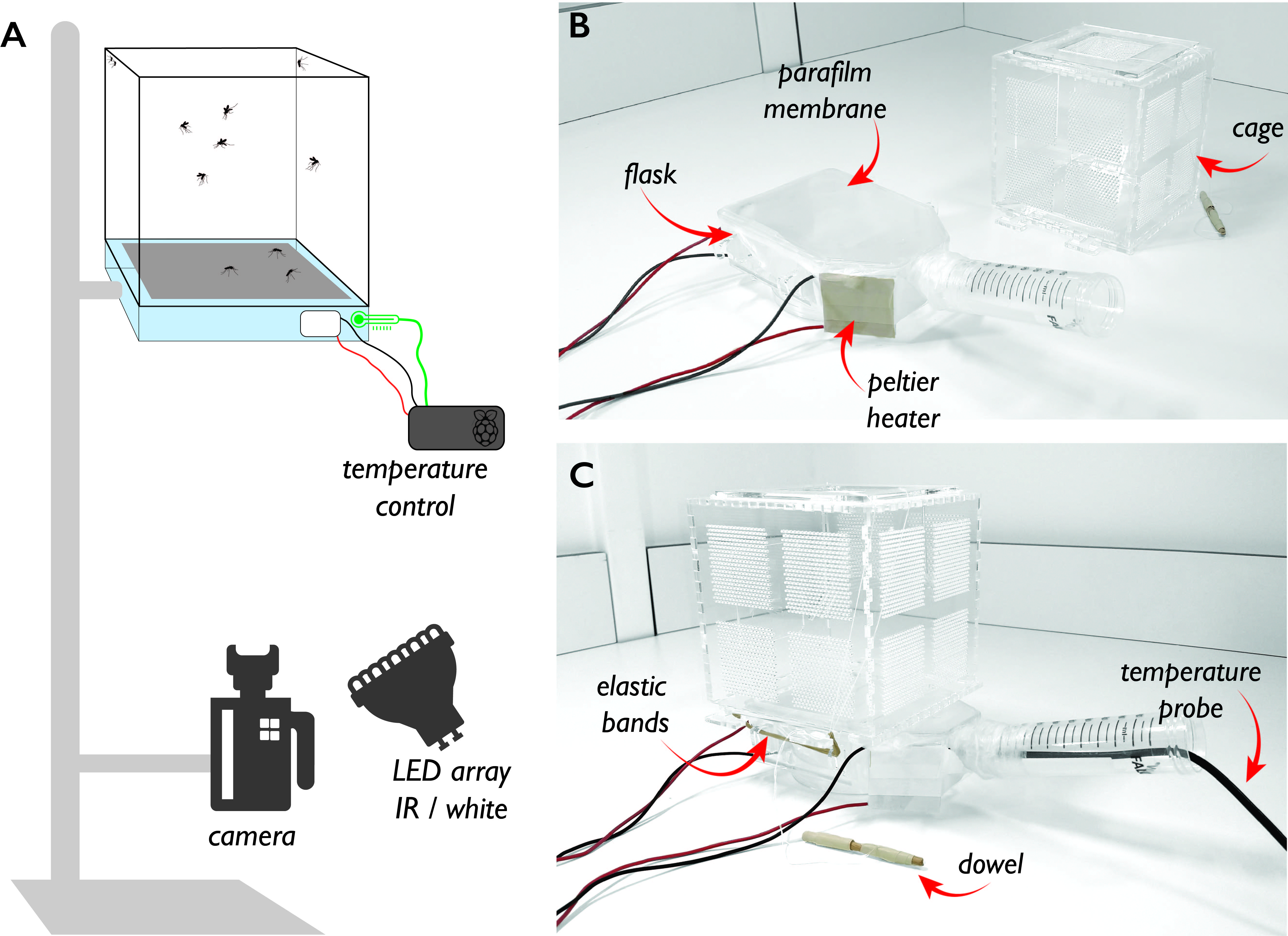}
\caption{The biteOscope setup. (\textbf{A}) Schematic of the setup, a camera is mounted under the bite substrate and cage to provide an abdominal view of mosquitoes interacting with the bite substrate. (\textbf{B}) Picture showing the bite substrate (including modified flask, Peltier elements, and Parafilm membrane) and cage separately. (\textbf{C}) Picture showing the bite substrate attached to the imaging window (bottom) of the cage using elastic bands. The temperature sensor is inserted into the water in the flask and the trap door is pulled up using the dowel. 
}
\label{fig:setup}
\end{figure}

\subsection{Cage fabrication}
Transparent cages can be easily constructed from clear cast acrylic using a laser cutter (e.g. a Trotec speedy 500, or epilog mini). In brief, laser cut 3mm transparent clear cast acrylic into the shapes provided in the supplied cage design files (Supplementary File 1). These pieces slot together to form a single transparent cube, see \FIG{setup}. The seams are crenulated for secure joining and if required may be permanently joined by application of dichloromethane. 

The upper surface of the cage is covered with a cotton mesh, glued in place with  non-toxic, non-smelling adhesive (e.g. a hot glue gun), before a further layer of acrylic is placed on top, glued and sealed around the edges. The two pieces of acrylic which make up the trap door may now be looped together with strong thread, through the precut holes. The 15 cm lengths of thread may then be threaded through the mesh in the upper surface before being tied and taped to a small dowel, to act as a handle for pulling open the trap door.

For added security against mosquitoes escaping, below the closed trap door a single-piece baseplate of acrylic is placed and secured using rubber bands around corresponding T-shaped tabs on the side of the cage and baseplate. Lastly, the small port on the side of the cage may be secured closed with tape once the mosquitoes have been transferred to the cage. 

\subsection{Creating the temperature controller}
\begin{figure}
\includegraphics[width=0.8\linewidth]{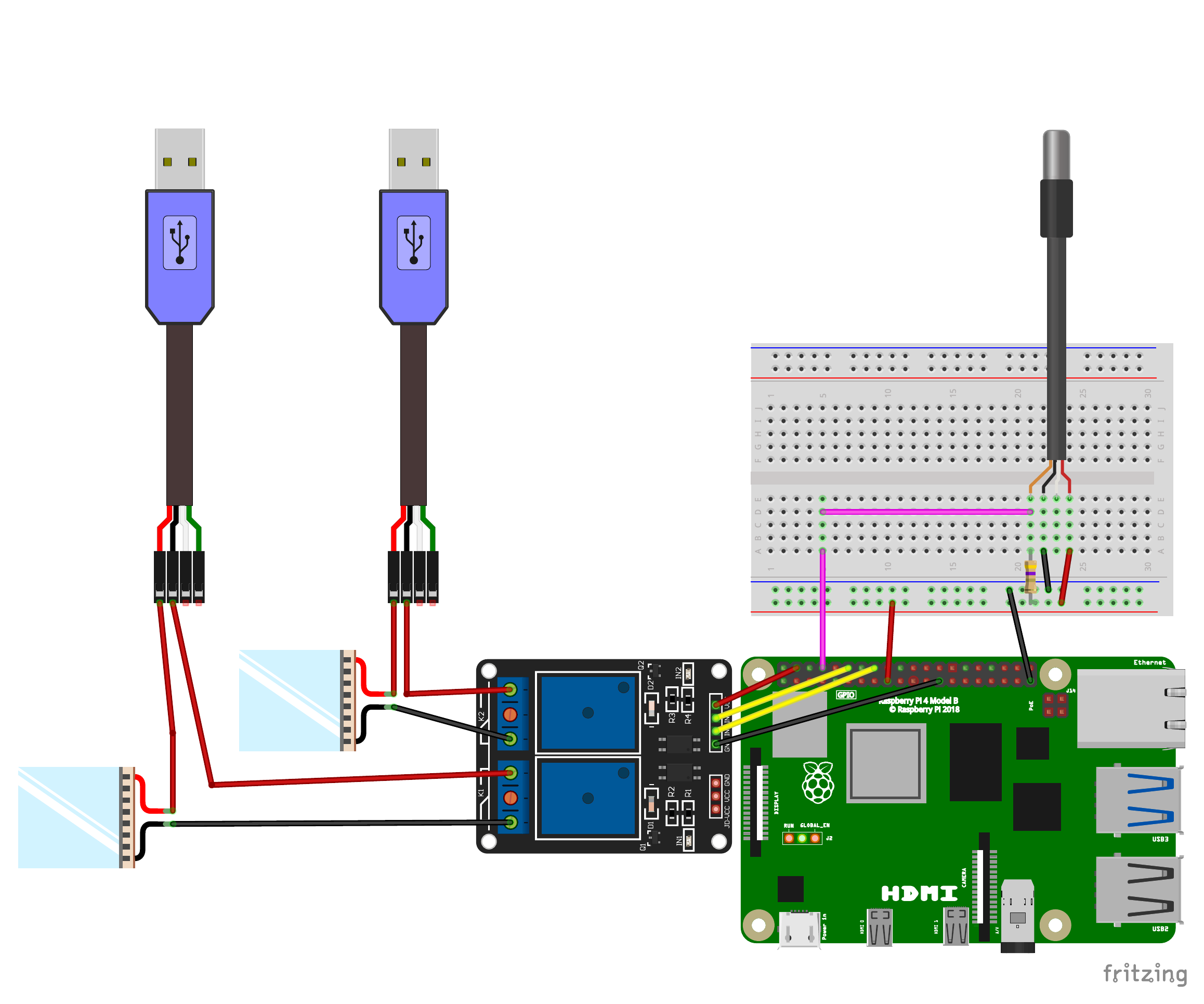}
\caption{Wiring scheme for the temperature controller. The temperature controller consists of a Raspberry Pi (green board), a relay switch (black), a waterproof DS18b20 temperature sensor, and two Peltier elements. Two USB plugs provide power to the Peltier elements. 
}
\label{fig:tempCont}
\end{figure}

The temperature controller is based on a Raspberry Pi (a small single board computer) and uses a waterproof 1-wire digital temperature sensor (DS18b20), a relay switch, and Peltier heating elements to maintain the bite substrate at the desired temperature. The wiring scheme for the temperature controller is shown in \FIG{tempCont}. Standard USB power adapters (e.g. a phone charger) can be used to provide power to the Peltier heating elements. 

To use the temperature sensor as an input source for the Raspberry Pi, the 1-wire interface has to be activated by running \verb+sudo raspi-config+, following instructions, and rebooting the Pi. This operation only has to be carried out once. Next, the \mintinline{python}{temperatureControl.py} script available on the \href{https://github.com/felixhol/biteOscope}{biteOscope github repository} should be transferred to the Raspberry Pi. Make sure that the Python script addresses the correct GPIO pins that  are connected to the relay input (the yellow wires in \FIG{tempCont}). 

\subsection{Bite substrate assembly}
The bottom opening of the cage, covered by the trap door, is designed to match the dimensions of the top surface of a 250 mL (75 cm\textsuperscript{2}) culture flask. This flask should be filled with hot, degassed water, so it will act as a transparent heating surface for the bite substrate. Degassing the water prevents bubble formation, improving image quality. The flask should be modified by the addition of a falcon tube to the opening, the falcon tube being cut open on the bottom and attached to the flask, creating an elongated neck through which a temperature sensor can be added. 

The flask will be attached to the underside of the acrylic cage by fastening the T-shaped tabs on the bottom of the cage to corresponding hooks on the side of the flask (hooks can be laser cut from acrylic and glued on the sides of the culture flask). During filming, the flask will be supported by clamp stands attached to the front and rear of the culture flask. The cage then rests on top of the flask and is secured using elastics bands between the T-shaped tabs and hooks. 

The water inside the culture flask is maintained at \SI{37}{\degreeCelsius} by two Peltier heating elements taped to the side of the flask. The heating elements are controlled by a Raspberry Pi and associated waterproof temperature probe, inserted into the neck of the flask. See \FIG{tempCont} for instructions regarding temperature controller assembly. For ease of use, connect the Peltier elements using alligator clips such that they can be easily connected and disconnected before and after the experiment. 

\section{Experimental Procedure}

Mosquitoes should be maintained under controlled insectary conditions e.g. \SI{28}{\degreeCelsius} and 70\% relative humidity. Depending on the question, female mosquitoes can be assayed between 6--25 days old. Experiments using day-biting mosquitoes (e.g. \emph{Aedes}) should be performed during daylight hours, while experiments with nocturnal species (e.g. \emph{Anopheles}) should be performed  in the dark during hours of peak biting activity. 

Depending on the question addressed, 1--35 mosquitoes can be assayed per cage. Depending on the response rate, more than 35 mosquitoes may lead to overcrowding of the field of view. Activity is typically highest in the first 15--30 min of an experiment, depending on the research question, multiple short experiments may therefore yield more data than a single long experiment. Triplicate cages (at minimum) are typically used for each condition. 

\subsubsection{1.} Deprive female mosquitoes of 10\% sucrose solution for >4 hours prior to the recording.
If using a large number of cages, the duration of sucrose deprivation and the time of day should be considered. Depending on the species of mosquito, the starvation period may need to be adjusted.

\subsubsection{2.} Fill the 250 mL culture flask with warm water and degas.
 
\indent \textit{If using multiple cages, a matching number of 250 mL flasks should be used.}

\subsubsection{3.} Prepare the saline-ATP solution containing 110mM NaCl, 20mM NaHCO\textsubscript{3}, and 1 mM ATP.\\

\subsubsection{4.} Assembling the bite substrate can be done in two ways: first applying the artificial meal to the culture flask and subsequently covering it with the membrane (step 4a), or vice versa (step 4b). 

\subsubsection{4a.} Apply approximately 5 mL saline-ATP solution to the rectangular outer surface of the culture flask. Carefully stretch the Parafilm membrane over the rectangular section of the tissue-culture flask containing the saline solution. The Parafilm and the outer surface of the flask create a thin fluid cell containing the artificial meal.

\indent \textit{Wear gloves at all times while manipulating the Parafilm and flask to avoid contamination.}

\subsubsection{4b.} Carefully stretch the Parafilm membrane over the rectangular section (outer surface) of the tissue-culture flask. Inject approximately 5 mL of the saline-ATP solution under the Parafilm with a syringe. The site of injection should be towards the neck of the flask to avoid damaging the parafilm within the behavioural space.

\indent \textit{Wear gloves at all times while manipulating the Parafilm and flask to avoid contamination.}

\subsubsection{5.} Set-up the behavioural arena: Place the cage on the bite substrate. Secure the cage to the bite substrate by attaching the elastic bands to the matching hooks and tabs on the flask and cage respectively. Attach the Peltier elements to the sides of the flask using adhesive tape and connect the wires of the Peltier elements to the power supplies. 

\subsubsection{6.} Insert the temperature probe into the flask (outside the field of view) and start the temperature control Python script \mintinline{python}{tempControlDual.py} on the Raspberry Pi.
     	
\indent \textit{Check the temperature of the water inside the flask prior to start.}

\subsubsection{7.} Position the behavioural arena above the camera so that the entire exposed bite surface is within the field of view. Connect the camera to the recording computer, start the recording software to preview the image. Position the LED arrays, if using infrared LEDs to record in the dark, attach an IR filter to the camera lens. Ensure the illumination is even across the bite surface. Adjust exposure time and or illumination intensity as necessary.  Ensure that the bite substrate is in focus.

\subsubsection{8.} Set the appropriate frame rate (typically 25 fps), exposure time, and write location. Start the recording.

\subsubsection{9.} Open the trap door to provide the mosquitoes access to the biting substrate.
     
\indent \textit{Check the focus and illumination of the recording.}

\subsubsection{10.} Provide a 20 second puff of CO\textsubscript{2} in the cage. 

\indent \textit{Depending on the experimental design, additional puffs of CO\textsubscript{2} may be applied in 10 min intervals.}

\section{Data analysis}

The acquired data can be analysed using a variety of methods. Below we provide two distinct computational pipelines: 1) a convolutional neural network (CNN trained using the DeepLabCut framework \citep{Mathis2018}) for mosquito and body part detection, and 2) tracking using `conventional' image processing techniques. The deep learning based method typically results in more accurate tracking and provides richer data, yet it also is computationally more intensive. While not strictly necessary, a dedicated GPU is recommended for the deep learning based approach. Cloud-based computing resources are an attractive resource when a GPU workstation is not available in-house. Imaging data can be acquired as a sequence of images (e.g. .tiff or .png files) or as a video (e.g. .mp4), these formats can be easily converted in both directions. The native format for the DeepLabCut framework is video, whereas image sequences are the most convenient input for the conventional approach.

\begin{figure}[t]
\includegraphics[width=\linewidth]{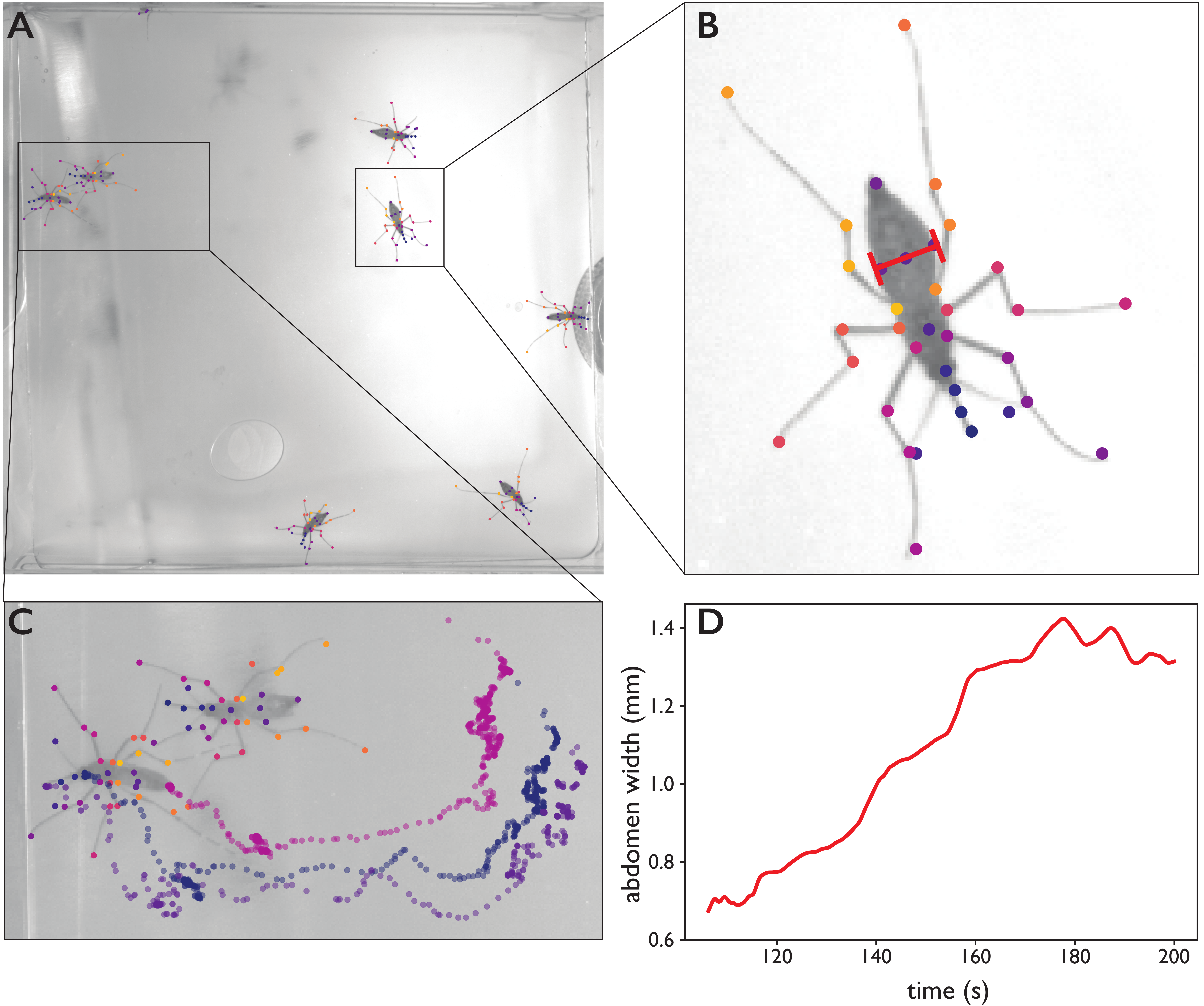}
\caption{Analysis of biteOscope data. (\textbf{A}) Frame showing an 8 x 8 cm field of view in which several \emph{Ae. aegypti} mosquitoes are visible. Dots indicate DLC-based detection of 38 key points on the mosquito body. (\textbf{B}) Magnification of an engorged mosquito shown in (A), dots show accurate detection of all key points. (\textbf{C}) Magnification of area shown in (A). Dots on the mosquito bodies show key point detection of a single frame, trails show the motion of the proboscis (purple), head (blue), and abdominal tip (magenta) of the left most mosquito for the preceding 28 seconds. (\textbf{D}) The width of the abdomen (red bar in (B)) can be used to assess engorgement status. During engorgement the width of the abdomen approximately doubles. 
}
\label{fig:tracking}
\end{figure}

\subsection{Deep learning based processing}
A pre-trained model to detect 38 key points on the body of \emph{Ae. aegypti} is available on github, see \FIG{tracking}. This ResNet50-based CNN was trained using DeepLabCut (DLC) and can be used for body part detection and tracking of up to approximately 18 \emph{Ae. aegypti} freely behaving in a 8 x 8 cm field of view. Alternatively, by labeling a modest amount of images this CNN can be re-trained to detect the body parts of mosquitoes having a different anatomy. This code is written in Python 3 and has been tested with DLC 2.2b8 on Ubuntu 18.04. 

\subsubsection{1.}
Download the biteOscope project from \href{https://github.com/felixhol/biteOscope}{Github}, and change the \mintinline{python}{project_path} variable in the \mintinline{python}{config.yaml} file to point to your local copy of the project: 

\begin{minted}{python}
import os
import glob
import deeplabcut

config_path = '/PATH_TO_CONFIG/config.yaml'
\end{minted}

\subsubsection{2.}
The biteOscope DLC project is set up to analyse mp4 videos using the DLC functions in the script below. This will first predict all body parts per frame and create a validation video showing all detections:

\begin{minted}[breaklines]{python}
videos = glob.glob('/PATH_TO_VIDEOS/*.mp4')
deeplabcut.analyse_videos(config_path, videos, shuffle=1, batchsize=8, dynamic=(True, 0.5, 100))
deeplabcut.create_video_with_all_detections(config_path, videos, 'DLC_resnet50_aedesNov16shuffle1_80000')
\end{minted}

\emph{Check if the accuracy of detected body parts is satisfactory.}

\subsubsection{3.}
Next, the detected body parts are assembled into animals, and animals are tracked across frames (resulting in `tracklets'). This creates a multilevel dataframe holding the predicted body part x,y coordinates of individual animals and their associated likelihood scores.

\begin{minted}[breaklines]{python}
deeplabcut.convert_detections2tracklets(config_path, videos, shuffle=1, videotype='mp4', trainingsetindex=0, track_method='box')
trackletPickles = glob.glob('/PATH_TO_VIDEOS/*_bx.pickle')

for tracklet in trackletPickles:
	deeplabcut.convert_raw_tracks_to_h5(config_path, tracklet)

deeplabcut.create_labeled_video(config_path, videos, videotype='.mp4', track_method='box', color_by='individual')
\end{minted}

\emph{Check tracking: all visible body parts should be labeled, individuals should be labeled in a single colour.}

\subsubsection{4.}
During an experiment mosquitoes may move in and out of the field of view. When e.g. mosquito \#1 moves out of the field of view ID `\#1' becomes vacant and DLC may re-assign `\#1' to the next mosquito entering the field of view. This is unwanted, as individual identities cannot be correctly assigned when individuals enter and leave the field of view. It is therefore more accurate to assign a new ID when individuals enter the field of view. The  \mintinline{python}{reID.py} code provided assigns DLC generated tracklets to unique IDs, the resulting dataframe can be used for downstream analysis. The biteOscope github repository contains Python code to calculate metrics like speed, activity, abdominal dilation, and feeding status per tracklet.

\begin{minted}[]{python}
DLC_output = '/PATH_TO_OUPUT/outputfile_bx.h5')
Dataframe = pd.read_hdf(DLC_output)
rD = reID(Dataframe)
\end{minted}

\subsection{Data processing using conventional image analysis}
Alternatively, images can be processed using conventional image processing techniques consisting of background subtraction, morphological operations, followed by blob detection. The biteOscope github repository contains code to perform these operations resulting in a dataframe holding the x,y coordinates of the center of mass of all mosquitoes. These data can be used to further process the data of individual mosquitoes. 

\subsubsection{1.}
Set all parameters in the \mintinline{python}{trackMosqNB.ipynb}.

\emph{Several parameters, including} \mintinline{python}{mTreshold} \emph{and} \mintinline{python}{searchRadius} \emph{may need to be adjusted.}

\subsubsection{2.}
Create a BG image used for background subtraction.

\emph{Check the appearance of the BG image, make sure inactive mosquitoes do not appear in the image.}

\subsubsection{3.}
Test mosquito detection with several frames, if detection is unsatisfactory, adjust \mintinline{python}{mTreshold} or the BG image.

\subsubsection{4.}
Detect centroids of mosquitoes in all frames and track detections.

\subsubsection{5.}
The created output can be used for downstream processing, e.g. using the \mintinline{python}{cropTracks_features.py} script provided on the github repository.

\bibliography{library}

\end{document}